\documentclass[twocolumn,amsmath,amssymb,floatfix,superscriptaddress,hyperref,reprint]{revtex4}
\usepackage[utf8]{inputenc}
\usepackage[T1]{fontenc}
\usepackage[usenames,dvipsnames]{xcolor}
\usepackage{graphicx}
\usepackage{dcolumn}  
\usepackage{amsbsy}  
\usepackage{ulem}
\usepackage{hyperref}

\allowdisplaybreaks[2] 

\usepackage{fourier}
\usepackage[scaled=0.875]{helvet}
\usepackage{courier}

\newcommand{\ind}[1]{_\mathrm{#1}}
\newcommand{\apref}[1]{appendix~\ref{#1}}

\newcommand{\transp}{^\mathrm{T}}
\DeclareMathOperator{\tr}{tr}

\newcommand{\dd}{{\mathrm{d}}}
\newcommand{\id}{{\mathrm{i}}}
\newcommand{\ed}{{\mathrm{e}}}

\newcommand{\zero}{{\boldsymbol{0}}}
\newcommand{\un}{{\boldsymbol{1}}}

\newcommand{\ff}{{\boldsymbol{f}}}
\renewcommand{\ggg}{{\boldsymbol{g}}}
\newcommand{\kk}{{\boldsymbol{k}}}
\newcommand{\pp}{{\boldsymbol{p}}}

\newcommand{\xx}{{\boldsymbol{x}}}
\newcommand{\yy}{{\boldsymbol{y}}}

\newcommand{\FF}{{\boldsymbol{F}}}
\newcommand{\GG}{{\boldsymbol{G}}}

\newcommand{\XXi}{{\boldsymbol{\Xi}}}
\newcommand{\eeta}{{\boldsymbol{\eta}}}
\newcommand{\nnabla}{{\boldsymbol{\nabla}}}
\newcommand{\ggamma}{{\boldsymbol{\gamma}}}

\begin{document}

\title{Diffusion of a particle quadratically coupled to a thermally fluctuating field}

\author{Vincent D\'emery}
\email{vincent.demery@polytechnique.edu}
\affiliation{Institut Jean Le Rond d'Alembert, CNRS and UPMC Universit\'e Paris 6, UMR 7190, F-75005 Paris, France, EU}

\begin{abstract}
We study the diffusion of a Brownian particle quadratically coupled to a thermally fluctuating field. In the weak coupling limit, a path-integral formulation allows to compute the effective diffusion coefficient in the cases of an active particle, that tends to suppress the field fluctuations, and of a passive particle, that only undergoes the field fluctuations. We show that the behavior is similar to what was previously found for a linear coupling: an active particle is always slowed down, whereas a passive particle is slowed down in a slow field and accelerated in a fast field. Numerical simulations show a good agreement with the analytical calculations. The examples of a membrane protein coupled to the curvature or composition of the membrane are discussed, with focus on the room for anomalous diffusion.
\end{abstract}

\maketitle

\section{Introduction}


Diffusion of an object interacting with its fluctuating environment has recently received a lot of attention: experimental studies have investigated the cases of colloidal beads diffusing along lipid bilayer tubes or through an actin network \cite{Wang2009}, insulin granules diffusing in $\beta$-cells \cite{Tabei2013}, or dielectric colloids subject to random optical forces generated by multiply scattered light \cite{Douglass2012}. A variety of behaviors have been observed, most of the time including anomalous diffusion, either in the mean squared displacement or in the probability distribution function. These observations call for a general theoretical framework able to describe diffusion in a complex environment. The considered systems can be cast into two classes:
if the object affects its environment, as in the first and second examples, it is called \emph{active}, if it does not, as in the third example, it is called \emph{passive}.

Among all the investigated systems, membrane proteins have concentrated much effort. One of the first theoretical studies is due to Saffman and Delbrück \cite{Saffman1975}, who have computed the hydrodynamic drag felt by a protein moving at constant velocity in a membrane; using the Einstein relation \cite{Einstein1905}, it allows to determine the diffusion coefficient. However, this calculation only gives a weak logarithmic dependence of the diffusion coefficient on the protein size that was contradicted later by accurate experiments \cite{Gambin2006}. Explaining the observed diffusion coefficient of membrane proteins has thus remained a major theoretical challenge. 
The numerous analytical and numerical investigations of the effective diffusion coefficient aim at taking into account two new effects: the geometrical penalty of free diffusion on a ruffled surface \cite{Reister2005, Reister_Gottfried2007, Naji2007, Leitenberger2008, Naji2009, Reister_Gottfried2010} and the interaction of the protein with a local parameter, most of the time the membrane curvature \cite{Reister2005, Leitenberger2008, Naji2009, Reister_Gottfried2010} but also its height \cite{Naji2007b}. 
These works have shown that the protein diffusion coefficient is reduced, whereas it is increased if the protein action on the membrane is neglected \cite{Naji2009, Reister_Gottfried2010}. This result is derived when the membrane equilibrates much faster than the protein moves; this is called the \emph{adiabatic} limit and it is the biologically relevant one  \cite{Reister_Gottfried2010}.

\begin{figure}
\begin{center}
\includegraphics[width=0.95\columnwidth]{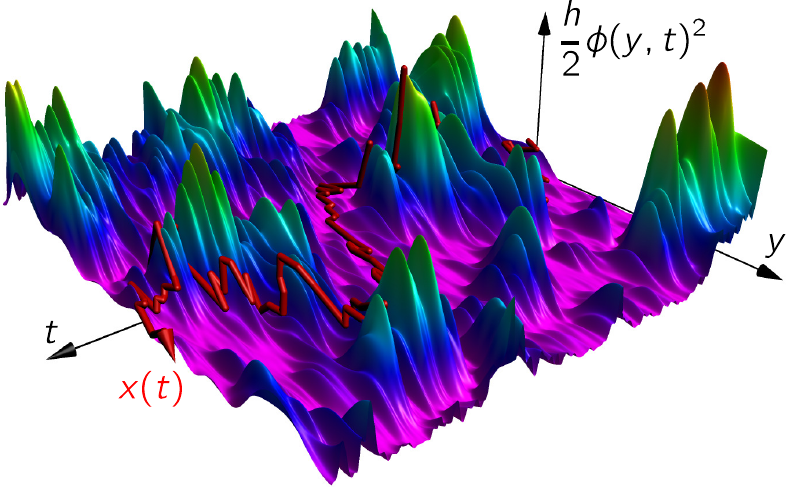}
\vspace{-5mm}
\end{center}
\caption{(Color online) Particle trajectory in a fluctuating field; it is quadratically coupled to the field and sees the effective potential $\frac{h}{2}\phi(\yy,t)^2$.}
\label{model}
\end{figure}

In previous studies \cite{Demery2010, Demery2010a, Dean2011, Demery2011}, we did a move towards a more general model, that aims at describing not only membrane proteins but also every kind of objects moving in a fluctuating environment, such as the ones presented sooner; a very generic picture is given in FIG~\ref{model}. In this model, a particle moving in a space of arbitrary dimension is linearly coupled to a scalar Gaussian field. We first computed the drag force felt by the object when it is pulled at constant velocity \cite{Demery2010,Demery2010a}. Naively, the Einstein relation would allow to deduce the diffusion coefficient but this is not the case since the drag is computed at \emph{constant velocity} whereas the Einstein relation needs the drag at \emph{constant force}. 
To clarify this issue, we computed the diffusion coefficient in the adiabatic limit, where the field is much faster than the particle, and showed that it can indeed be inferred from the drag coefficient computed at constant velocity \cite{Dean2011, Demery2011}. In this limit, we showed that an active particle is always slowed down by its coupling to the field, whereas a passive particle is always accelerated: this corresponds to what was observed for the motion of membrane proteins \cite{Naji2009, Reister_Gottfried2010}. However, in a more general theory one has to consider the non-adiabatic limit, where we showed that the previous statement breaks down:
in a slow field a passive particle is slowed down too \cite{Demery2011}.

This first generalization needs to be further extended in order to be able to model more couplings between the object and the environment. For instance, when a protein is coupled to membrane curvature, there is a superposition of two effects: the protein may impose a spontaneous curvature to the membrane, and, if it is stiffer than the membrane, it can also reduce the membrane fluctuations. Whereas the first effect is well described by a linear interaction between the protein and the membrane, the second one needs a quadratic interaction. 
What was proved for the linear interaction needs to be investigated for the quadratic interaction: this article fills this gap. The drag force at constant velocity has already been computed for the quadratic interaction \cite{Demery2011a}. 

In this article, we introduce the general model in section \ref{sec_model} and explain how it describes a protein coupled to membrane curvature or composition. 
Analytical calculations are performed in section \ref{sec_anal}. Firstly, we derive the exact path-integral representation of the particle motion. Secondly, we show that in the weak coupling limit it reduces to an effective dynamics for the particle, containing a memory term. Thirdly, we address the adiabatic limit, derive a Markovian dynamics and compute the effective diffusion coefficient; it appears that it can be deduced from the drag computed at constant velocity. 
Fourthly, we compute the effective diffusion coefficient for a weak coupling but outside the adiabatic limit and show that an active particle is always slowed down whereas a passive particle is slowed down in a slow field and accelerated in a fast field. 
These results for a quadratic coupling, albeit very similar to those for a linear coupling, are new; moreover, to get to the effective diffusion equation, the quadratic coupling requires new techniques that are detailed here.
Analytical results are compared to numerical simulations on a simple unidimensional system in section \ref{sec_num_sim}.
The possibility for anomalous diffusion is discussed in section \ref{sec_anomalous}.

\section{Model}\label{sec_model}

\subsection{General model}

Our general model is 
relatively simple but contains several adjustable parameters that allow to use it to model many different systems. Some of these ones are given as examples at the end of this section. 	

This model is close to the one presented in \cite{Demery2011}, with the exception that the coupling with the field is here quadratic in the field, making the interaction "fluctuations induced". This quadratic coupling has already been introduced in \cite{Demery2011a}, albeit a in less general formulation, where the particle was pulled at constant velocity; here the particle is submitted to a thermal noise.

We consider a particle located at $\xx$ in a $d$-dimensional space, with a quadratic coupling to a Gaussian field $\phi(\yy)$. The energy of the particle-field system is given by
\begin{equation}
H[\phi,\xx]= \frac{1}{2}\int \phi(\yy)[\Delta \phi](\yy) \dd\yy + \frac{h}{2}[K\phi](\xx)^2.
\end{equation}
The notations used for functional operators are defined in \apref{ap_operators}.
The first term is the quadratic energy of a free field, and the operator $\Delta$ gives the shape of the field we consider. The second term is the quadratic coupling between the field and the particle, the particle shape is defined by the operator $K$, and $h$ is the coupling constant. The pure Casimir interaction where the particle suppress the field fluctuations is obtained in the limit $h\rightarrow\infty$. The operators $\Delta$ and $K$ are translation invariant and isotropic (cf. \apref{ap_operators}).

We start with an overdamped Langevin equation for the whole system, the equilibrium of which is thus given by the Gibbs-Boltzmann statistics. The particle position evolves as follows,
\begin{align}
\dot\xx(t) & =-\kappa_x \frac{\delta H}{\delta \xx}[\phi(\yy,t),\xx(t)]+\sqrt{\kappa_x}\eeta(t),\\
& = -\frac{h\kappa_x}{2}\nnabla \left[ (K\phi)^2\right](\xx(t),t)+\sqrt{\kappa_x}\eeta(t). \label{dyna_part}
\end{align}
$\kappa_x$ is the mobility of the particle and $\eeta(t)$ is a Gaussian white noise with correlator
\begin{equation}\label{correl_eta}
\left\langle \eeta(t)\eeta(t')\transp \right\rangle=2T\delta(t-t')\un.
\end{equation}
The superscript "T" denotes the transposition and $\un$ is the identity matrix. The field evolution is 
\begin{align}
\dot\phi(\yy,t)&=-\kappa_\phi \left(R \frac{\delta H}{\delta\phi} \right)[\phi(\yy,t),\xx(t)]+\sqrt{\kappa_\phi}\left(\sqrt{R}\xi\right)(\yy,t),\\
& = -\kappa_\phi \left[(R\Delta\phi)(\yy,t)\right.\nonumber\\
&\quad\quad + \left. h(K\phi)(\xx(t),t)\left(R K\delta\right)(\yy-\xx(t)) \right]\nonumber \\
&\quad\quad + \sqrt{\kappa_\phi}\left(\sqrt{R}\xi\right)(\yy,t). \label{dyna_champ_bd}
\end{align}
$\kappa_\phi$ is the "mobility" (or evolution rate) of the field, $R$ is a translation-invariant isotropic operator defining the field dynamics, and $\xi(\yy,t)$ is a functional Gaussian white noise with correlator
\begin{equation}\label{correl_xi}
\left\langle \xi(\yy,t)\xi(\yy',t') \right\rangle=2T\delta(\yy-\yy')\delta(t-t').
\end{equation}
To take into account the case of a passive particle, i.e. a particle that does not affect the field, we introduce a \emph{feedback parameter} $\zeta$ in (\ref{dyna_champ_bd}) that becomes
\begin{align}
\dot\phi(\yy,t)& = -\kappa_\phi \left[(R\Delta\phi)(\yy,t)\right.\nonumber\\
&\quad\quad + \left.\zeta h(K\phi)(\xx(t),t)\left(R K\delta\right)(\yy-\xx(t)) \right]\nonumber \\
&\quad\quad + \sqrt{\kappa_\phi}\left(\sqrt{R}\xi\right)(\yy,t). \label{dyna_champ}
\end{align}
The previous equation is recovered with $\zeta=1$ whereas a completely passive particle corresponds to $\zeta=0$. Except for $\zeta=1$, the dynamics does not satisfy detailed balance, and the system is out of equilibrium. A trajectory of the particle-field system is given as an example in FIG.~\ref{model}.

In this paper, we will focus on the \emph{effective} or \emph{late-time} diffusion coefficient, defined by
\begin{equation}
D\ind{eff}=\lim_{t\rightarrow\infty} \frac{\left\langle [\xx(t)-\xx(0)]^2 \right\rangle}{2dt}.
\end{equation}
Of course, this number is only defined if the diffusion is normal, and we will restrict ourselves to this case here. However, it can give a hint on the kind of diffusion: $D\ind{eff}=0$ indicates subdiffusion and $D\ind{eff}=\infty$ signals superdiffusion. Without coupling, the bare diffusion coefficient is
\begin{equation}
D_x=T\kappa_x.
\end{equation}

\subsection{Examples}\label{ssec_model_examples}

Our model applies to many systems where the presence of an inclusion penalizes the fluctuations of an order parameter in its environment. In our examples the inclusion is a membrane protein that can be coupled to the membrane curvature or composition. The dimension of the space accessible to the protein is $d=2$.

The case of a protein affecting the curvature has been extensively discussed \cite{Goulian1993, Reister2005, Reister_Gottfried2007, Naji2007, Leitenberger2008, Naji2009, Reister_Gottfried2010, Bitbol2010}. First, we define the free field quantities: the membrane height corresponds to our field $\phi(\yy,t)$ and its free field energy is given by
\begin{equation}
H_0[\phi]=\frac{1}{2}\int \left[\kappa\ind{m}\left(\nnabla^2\phi(\yy)\right)^2+\sigma(\nnabla\phi(\yy))^2 \right]\dd\yy,
\end{equation}
where $\kappa\ind{m}$ is the membrane bending modulus and $\sigma$ its surface tension. 
In our formalism, the corresponding $\Delta$ operator is given in Fourier space by
\begin{equation}
\tilde \Delta(\kk)=\kappa\ind{m}\kk^2\left(\kk^2+m^2\right),
\end{equation}
where $m=\sqrt{\sigma/\kappa\ind{m}}$ is the inverse of the correlation length $l$. The protein tries to impose its spontaneous curvature, giving rise to the coupling
\begin{equation}
H\ind{int}[\phi,\xx]=\frac{\kappa\ind{p}}{2}\left[\nnabla^2\phi(\xx)-2C\ind{p}  \right]^2-\frac{\kappa\ind{m}}{2}\left[\nnabla^2\phi(\xx)\right]^2,
\end{equation}
where $C\ind{p}$ is the spontaneous protein curvature and $\kappa\ind{p}$ its bending rigidity. Although in most of the studies the interaction has a non-trivial shape, we need to consider it point-like here. The size of the protein can be introduced as an effective cut-off at the end of the computation. This interaction term can be decomposed into a linear contribution, that tends to deform the membrane, plus a quadratic contribution, that tends to suppress its fluctuations. The case of a linear interaction has been treated in \cite{Demery2011}, and here we are only interested in the quadratic term. Physically, this corresponds to a tough, but flat protein: $\Delta\kappa=\kappa\ind{p}-\kappa\ind{m}>0$, $C\ind{p}=0$, and the interaction is defined by
\begin{equation}
h=\Delta\kappa,\quad \tilde K(\kk)=\kk^2.
\end{equation}
Finally, the membrane dynamics is given by the Oseen hydrodynamic tensor \cite{Lin2004},
\begin{equation}
\tilde R(\kk)=\frac{1}{4\eta|\kk|},
\end{equation}
where $\eta$ is the viscosity of the surrounding fluid. The effect of projection due to the diffusion on a ruffled surface \cite{Reister2005, Reister_Gottfried2007, Naji2007, Leitenberger2008, Naji2009, Reister_Gottfried2010} is neglected here.


Lipid membranes are made of several kinds of lipids, and proteins can also be coupled to the composition field $\phi$ \cite{Reynwar2008}, that fluctuates on growing length scales as the system is tuned close to the miscibility transition \cite{Honerkamp_Smith2009}. The composition field is ruled by the following operators \cite{Reynwar2008}
\begin{align}
\tilde\Delta(\kk)&=\kk^2+m^2,\\
\tilde K(\kk) & = 1,\\
\tilde R(\kk) & = \kk^2.
\end{align}
The operator $R$ ensures that the average composition is conserved. Again, $m$ is the inverse correlation length, and goes to zero close to the miscibility transition as
\begin{equation}\label{mass_temp}
m=l^{-1}\sim (1-T/T\ind{c})^\nu,
\end{equation}
where $T\ind{c}$ is the miscibility temperature and $\nu=1$ is the exponent observed experimentally on Giant Unilamellar Vesicles \cite{Honerkamp_Smith2009}. The interaction between two proteins has been calculated exactly at the critical point in \cite{Machta2012}. 

Proteins can be coupled to other membrane fields such as the local membrane thickness \cite{Naji2007b}. These parameters can be advected, or not, by the flow created by the protein motion. This effect, discussed in \cite{Bitbol2011,Camley2012}, is neglected here.

\section{Analytical calculations}\label{sec_anal}

\subsection{Path-integral formalism and effective action}\label{}

The evolution of the particle-field system, that is completely defined by (\ref{dyna_part}, \ref{correl_eta}, \ref{correl_xi}, \ref{dyna_champ}), can be cast in a path-integral formulation. Following the procedure used in \cite{Demery2011,Aron2010} with the It\=o convention, the partition function for the whole system is of the form
\begin{align}
Z=&\int \exp \left(-S_0[\xx,\pp]-S_0^\phi[\phi,\psi]-S\ind{int}[\xx,\pp,\phi,\psi] \right)\nonumber\\
&\quad\quad\times[\dd\xx][\dd\pp][\dd\phi][\dd\psi].
\end{align}
$\pp(t)$ and $\psi(\yy,t)$ are the response fields \cite{Aron2010} associated respectively with $\xx(t)$ and $\phi(\yy,t)$, and the action contains three terms. $S_0[\xx,\pp]$ is the action of the pure Brownian motion (i.e. without interaction with the field) and reads \cite{Demery2011}
\begin{equation}
S_0[\xx,\pp]=-\id \int\pp(t)\cdot\dot\xx(t)\dd t+D_x\int\pp(t)^2\dd t.
\end{equation}
The free field action is of the same form:
\begin{align}\label{act_champ_libre}
S_0^\phi[\phi,\psi]=& -\id\int\psi(\yy,t)\left[\dot\phi(\yy,t)+\kappa_\phi(R\Delta\phi)(\yy,t) \right]\dd \yy\dd t\nonumber \\
& + T\kappa_\phi\int\psi(\yy,t)(R\psi)(\yy,t)\dd \yy\dd t.
\end{align}
Finally, the interaction action couples the particle and the field:
\begin{align}
S\ind{int}[\xx,\pp,\phi,\psi]=& -\id \frac{h\kappa_x}{2}\int \pp(t)\cdot \nnabla \left([K\phi]^2 \right)(\xx(t),t)\dd t\nonumber\\
&-\id \zeta h\kappa_\phi\int(K\phi)(\xx(t),t)\left(KR\psi\right)(\xx(t),t)\dd t.
\end{align}

To get an effective action $S\ind{int}^\mathrm{eff}$ for the particle, we integrate over $(\phi,\psi)$:
\begin{multline}
\exp \left(-S\ind{int}^\mathrm{eff}[\xx,\pp] \right)=\\ \int \exp \left(-S_0^\phi[\phi,\psi]-S\ind{int}[\xx,\pp,\phi,\psi] \right)[\dd\phi][\dd\psi].
\end{multline}
This integral is quadratic and can be computed. We need the following formula: for a random Gaussian vector $X$ and a matrix $A$, 
\begin{align}
\left\langle \exp \left(X\transp AX \right) \right\rangle_X&=\det \left(\un-2A \left\langle XX\transp \right\rangle \right)^{-1/2}\\
&=\exp \left[-\frac{1}{2}\tr \left(\log \left[\un- 2A \left\langle XX\transp \right\rangle\right] \right) \right].\label{moy_gauss}
\end{align}
We use it considering $(\phi,\psi)$ as a Gaussian random variable weighted by $S_0^\phi$, the interaction action $S\ind{int}[\xx,\pp,\phi,\psi]$  playing the role of $X\transp A X$. 
Fourier transforming the fields in space allows us to write the interaction action
\begin{multline}
S\ind{int}[\xx,\pp,\phi,\psi]= \int \left[\tilde A_{\phi\phi}(-\kk,-\kk',t,t')\tilde\phi(\kk,t)\tilde\phi(\kk',t') \right.\\
 +\left.\tilde A_{\phi\psi}(-\kk,-\kk',t,t')\tilde\phi(\kk,t)\tilde\psi(\kk',t')\right]\frac{\dd\kk\dd\kk'}{(2\pi)^{2d}}\dd t\dd t',
\end{multline}
where
\begin{align}
\tilde A_{\phi\phi}(\kk,\kk',t,t')&=-\frac{h\kappa_x}{2}\pp(t)\cdot(\kk+\kk')\tilde K(\kk)\tilde K(\kk')\nonumber\\
&\quad\quad\quad\times \ed^{-\id(\kk+\kk')\cdot\xx(t)}\delta(t-t'),\\
\tilde A_{\phi\psi}(\kk,\kk',t,t')&=-\id\zeta h\kappa_\phi\tilde K(\kk)\tilde K(\kk')\tilde R (\kk')\ed^{-\id(\kk+\kk')\cdot\xx(t)}\delta(t-t').
\end{align}
We then need the free field correlation functions that are computed in appendix \ref{ap_correl_champ}:
\begin{align}
\tilde C_{\phi\phi}(\kk,\kk',t,t') & = \left\langle \tilde\phi(\kk,t)\tilde\phi(\kk',t') \right\rangle\\
& = \frac{T}{\tilde\Delta(\kk)}\ed^{-\kappa_\phi|t-t'|\tilde R(\kk)\tilde\Delta(\kk)}(2\pi)^d\delta(\kk+\kk'),\\
\tilde C_{\phi\psi}(\kk,\kk',t,t') & = \left\langle \tilde\phi(\kk,t)\tilde\psi(\kk',t') \right\rangle\\
& = \id\ed^{-\kappa_\phi|t-t'|\tilde R(\kk)\tilde\Delta(\kk)}\theta(t-t')(2\pi)^d\delta(\kk+\kk').
\end{align}
$\theta(t)$ is the Heavyside function, with $\theta(0)=0$ since we use the It\=o convention.
We introduce the operator
\begin{equation}
M=A_{\phi\phi}C_{\phi\phi}+A_{\phi\psi}C_{\psi\phi},
\end{equation}
whose Fourier transform reads
\begin{multline}
\tilde M(\kk,\kk',t,t')=h\ed^{-\id (\kk+\kk')\cdot\xx(t)-\kappa_\phi|t-t'|\tilde R(\kk')\tilde\Delta(\kk')} \tilde K(\kk)\tilde K(\kk')\\ \times\left[-\frac{T\kappa_x}{2}\frac{\pp(t)\cdot(\kk+\kk')}{\tilde\Delta(\kk')}+\zeta\kappa_\phi \tilde R(\kk')\theta(t'-t)\right].
\end{multline}
The formula  (\ref{moy_gauss}) gives the effective interaction action as
\begin{equation}\label{act_eff}
S\ind{int}^\mathrm{eff}[\xx,\pp]=\frac{1}{2}\tr \left[\log \left(\un+2M[\xx,\pp] \right) \right];
\end{equation}
it allows us to write the partition function
\begin{equation}
Z=\int \exp \left(-S_0[\xx,\pp]-S\ind{int}^\mathrm{eff}[\xx,\pp] \right)[\dd\xx][\dd\pp].
\end{equation}
This expression is exact and constitutes our first new result.

The logarithm appearing in the interaction action (\ref{act_eff}) makes it very difficult to handle; its signification is discussed later. To go further, we have to expand the logarithm; this can be done for a small coupling $h\ll 1$.

\subsection{Effective particle dynamics}\label{}

In the weak interaction limit ($h\ll 1$) a simpler expression can be obtained by expanding the effective interaction action (\ref{act_eff}) in powers of the coupling constant $h$. 
We start by developing the logarithm it contains in powers of $M$, that is proportional to $h$:
\begin{equation}
\log(\un+2M)=2M-2M^2+\mathcal{O}\left(h^3\right).
\end{equation}
Now, the trace can be taken explicitly:
\begin{equation}
\tr(M) = \int \tilde M(\kk,-\kk,t,t)\frac{\dd\kk}{(2\pi)^d}\dd t=0,
\end{equation}
since $\theta(0)=0$ with our convention. The fact that the first term, proportional to $h$, does not contribute is not surprising: to feel the field, the particle must interact with it at least twice. For the second term, we need
\begin{align}
\tr \left(M^2 \right) & = \int \tilde M(\kk,\kk',t,t')\tilde M(-\kk',-\kk,t',t)\frac{\dd\kk\dd\kk'}{(2\pi)^{2d}}\dd t\dd t'\nonumber\\
& = h^2T\kappa_x\int \ed^{\id(\kk+\kk')\cdot\left[\xx(t)-\xx(t')\right]-\kappa_\phi\left[\tilde R(\kk)\tilde\Delta(\kk)+\tilde R(\kk')\tilde\Delta(\kk')\right]|t-t'|}\nonumber\\
&\quad\times\left(\zeta\kappa_\phi\pp(t)\cdot(\kk+\kk') \frac{\tilde R(\kk)}{\tilde\Delta(\kk')}\theta(t-t') \right.\nonumber\\
&\quad\quad -\left. \frac{T\kappa_x}{4}\frac{\pp(t)\cdot(\kk+\kk')\pp(t')\cdot(\kk+\kk')}{\tilde\Delta(\kk)\tilde\Delta(\kk')}\right)\nonumber\\
&\quad\times \tilde K(\kk)^2\tilde K(\kk')^2 \frac{\dd\kk\dd\kk'}{(2\pi)^{2d}}\dd t\dd t'.
\end{align}
This expression has been obtained after several variable changes on the different terms. The action at the order $h^2$ is thus of the form
\begin{align}
S\ind{int,2}^\mathrm{eff}[\xx,\pp]&=-\tr \left(M^2 \right)\nonumber\\
&=\id\int_{t>t'}\pp(t)\cdot\FF(\xx(t)-\xx(t'),t-t')\dd t\dd t'\nonumber\\
&\quad+\frac{T}{2}\int\pp(t)\transp\GG(\xx(t)-\xx(t'),t-t')\pp(t')\dd t\dd t',\label{act_int_2}
\end{align}
where
\begin{multline}
\FF(\xx,t)=\id\zeta h^2T\kappa_\phi\kappa_x \int \left(\kk+\kk' \right)\frac{\tilde K(\kk)^2\tilde K(\kk')^2\tilde R(\kk)}{\tilde\Delta(\kk')} \\
\quad\times \ed^{\id(\kk+\kk')\cdot\left[\xx(t)-\xx(t')\right]-\kappa_\phi\left[\tilde R(\kk)\tilde\Delta(\kk)+\tilde R(\kk')\tilde\Delta(\kk')\right]|t-t'|}\frac{\dd\kk\dd\kk'}{(2\pi)^{2d}}.
\end{multline}
and
\begin{multline}
\GG(\xx,t)=\frac{h^2T\kappa_x^2}{2}\int (\kk+\kk')(\kk+\kk')\transp \frac{\tilde K(\kk)^2\tilde K(\kk')^2}{\tilde\Delta(\kk)\tilde\Delta(\kk')}\\
\quad\times \ed^{\id(\kk+\kk')\cdot\left[\xx(t)-\xx(t')\right]-\kappa_\phi\left[\tilde R(\kk)\tilde\Delta(\kk)+\tilde R(\kk')\tilde\Delta(\kk')\right]|t-t'|}  \frac{\dd\kk\dd\kk'}{(2\pi)^{2d}}.
\end{multline}

In the truncated interaction action (\ref{act_int_2}), we recognize the action associated to the effective evolution equation \cite{Demery2011}
\begin{multline}\label{evol_eff}
\dot\xx(t)=\sqrt{\kappa_x}\eeta(t)\\+\int_{-\infty}^t\FF(\xx(t)-\xx(t'),t-t')\dd t'+\XXi(\xx(t),t),
\end{multline}
where the noise $\XXi(\xx,t)$ has a correlator
\begin{equation}
\left\langle \XXi(\xx(t),t)\XXi(\xx(t'),t')\transp \right\rangle=T\GG(\xx(t)-\xx(t'),t-t').
\end{equation}

This dynamics is of the same form as the one obtained in \cite{Demery2011} for a linear coupling between the field and the particle. The case of a linear coupling being simpler, the effective dynamics is exact, whereas it is a weak coupling expansion here. This dynamics reveals a two-times interaction with the field, at times $t'$ and $t$. This is also the sense of the $h$-expansion performed in the effective action (\ref{act_eff}) to get the effective dynamics: to restrict ourselves to two-times interactions. Expanding further the logarithm in this action would add more terms to the effective dynamics, corresponding to four-times interactions, six-times interactions, etc.


\subsection{Adiabatic limit}\label{sub_adia}

In the adiabatic limit, where the field is much faster than the particle ($\kappa_x\ll\kappa_\phi$), the field memory time reduces to zero and the effective dynamics becomes Markovian.
We follow the procedure used in \cite{Dean2011} and compute $\FF$ and $\GG$ to the first order in $\kappa_x/\kappa_\phi$. For $\FF$, this amounts to write
\begin{multline}
\FF(\xx(t)-\xx(t'),t-t')=\\(t-t')\dot\xx(t)\cdot\nnabla\FF(\zero,t-t')+\mathcal{O}\left([t-t']^2 \right).
\end{multline}
Integrating over $t'$, we get to the first order in $\kappa_x/\kappa_\phi$
\begin{align}\label{F_adia}
&\int_{-\infty}^t\FF(\xx(t)-\xx(t'),t-t')\dd t'= -\frac{\kappa_x}{\kappa_\phi}\frac{\zeta h^2T}{2d} \nonumber\\
&\quad\times\left[\int \frac{\left(\kk^2+\kk'^2\right)\tilde K(\kk)^2\tilde K(\kk')^2}{\tilde\Delta(\kk)\tilde\Delta(\kk')[\tilde R(\kk)\tilde\Delta(\kk)+\tilde R(\kk')\tilde\Delta(\kk')]}\frac{\dd\kk\dd\kk'}{(2\pi)^{2d}} \right]\dot\xx(t).
\end{align}
For $\GG$, we write
\begin{equation}
\GG(\xx(t)-\xx(t'),t-t')=\GG(\zero,t-t')+\mathcal{O}(t-t'),
\end{equation}
and we integrate over $t'$ to get, again to the first order in $\kappa_x/\kappa_\phi$,
\begin{multline}\label{G_adia}
\GG(\xx(t)-\xx(t'),t-t')= \delta(t-t')\frac{\kappa_x}{\kappa_\phi}\frac{h^2T\kappa_x}{d}\\ \times \left[\int \frac{\left(\kk^2+\kk'^2\right)\tilde K(\kk)^2\tilde K(\kk')^2}{\tilde\Delta(\kk)\tilde\Delta(\kk')[\tilde R(\kk)\tilde\Delta(\kk)+\tilde R(\kk')\tilde\Delta(\kk')]}\frac{\dd\kk\dd\kk'}{(2\pi)^{2d}} \right].
\end{multline}
We recognize the same integral in (\ref{F_adia}) and (\ref{G_adia}), so we introduce
\begin{equation}
\lambda=\frac{h^2T}{2d} \int \frac{\left(\kk^2+\kk'^2\right)\tilde K(\kk)^2\tilde K(\kk')^2}{\tilde\Delta(\kk)\tilde\Delta(\kk')[\tilde R(\kk)\tilde\Delta(\kk)+\tilde R(\kk')\tilde\Delta(\kk')]}\frac{\dd\kk\dd\kk'}{(2\pi)^{2d}},
\end{equation}
that allows us to write the effective evolution equation for the particle, to the order $h^2$ in the coupling and $\kappa_x/\kappa_\phi$ in the evolution rates,
\begin{equation}
\dot\xx(t)=-\zeta\frac{\kappa_x}{\kappa_\phi}\lambda\dot\xx(t)+\sqrt{\kappa_x}\eeta(t)+\sqrt{\frac{\kappa_x^2}{\kappa_\phi}\lambda}\eeta'(t),
\end{equation}
where $\eeta'(t)$ is a Gaussian white noise with the same correlator as $\eeta(t)$ (\ref{correl_eta}). 

We recognize the drag coefficient found in \cite{Demery2011a}. This result is very similar to the case of a linear coupling \cite{Dean2011}: in the adiabatic limit, the drag coefficient allows to derive a Markovian effective dynamics; we have shown here that this property also holds for a quadratic interaction, and may thus be universal.


With this dynamics, the effective diffusion coefficient is easy to compute:
\begin{equation}
D\ind{eff}^\mathrm{adia}=D_x \left(1-(2\zeta-1)\frac{\kappa_x}{\kappa_\phi}\lambda \right).
\end{equation}
This expression indicates that the diffusion coefficient is decreased for an active particle and increased for a passive one, and corresponds to what was found for proteins coupled to membrane curvature \cite{Naji2009,Reister_Gottfried2010}.
We show in the following that this is only true in the adiabatic limit.

\subsection{Diffusion coefficient outside the adiabatic limit}\label{}

The effective diffusion coefficient can also be computed outside the adiabatic limit; this was done for a linear coupling in \cite{Dean2011} for the active case and in \cite{Demery2011} for the general case. The path-integral method used in \cite{Demery2011} can be generalized and applied to the effective diffusion equation (\ref{evol_eff}).


First, we put the functions $\FF$ and $\GG$ in the general form
\begin{align}
\FF(\xx,t) &=\id\int\ed^{\id\ggamma(Q)\cdot\xx-\omega(Q)|t|}\ff(Q)\dd Q,\\
\GG(\xx,t) &=\int\ed^{\id\ggamma(Q)\cdot\xx-\omega(Q)|t|}\ggg(Q)\dd Q,
\end{align}
where in our case
\begin{align}
Q &= (\kk,\kk'),\\
\ggamma(Q) &=\kk+\kk',\\
\omega(Q) &=\kappa_\phi\left[\tilde R(\kk)\tilde\Delta(\kk)+\tilde R(\kk')\tilde\Delta(\kk')\right],\\
\ff(Q) &=\frac{h^2D_x\kappa_\phi}{(2\pi)^{2d}} \ggamma(Q)\frac{\tilde K(\kk)^2\tilde K(\kk')^2\tilde R(\kk)}{\tilde\Delta(\kk')},\\
\ggg(Q) &=\frac{h^2D_xT}{2(2\pi)^{2d}} \ggamma(Q)\ggamma(Q)\transp \frac{\tilde K(\kk)^2\tilde K(\kk')^2}{\tilde\Delta(\kk)\tilde\Delta(\kk')}.
\end{align}
In this form, the functions $\FF$ and $\GG$ are very similar to those found in \cite{Demery2011} for a linear coupling to the field. The treatment that leads to the effective diffusion coefficient is thus the same; we briefly recall it but refer the reader to \cite{Demery2011} for more details.

The average quadratic displacement between times 0 and $t$ is approximated to the order $h^2$ by
\begin{multline}
\left\langle \xx_0(t)^2 \right\rangle= \left\langle \xx_0(t)^2 \right\rangle_0 \left(1-\left\langle S_\mathrm{int,2}^\mathrm{eff}[\xx,\pp]  \right\rangle_0  \right)\\- \left\langle \xx_0(t)^2 S_\mathrm{int,2}^\mathrm{eff}[\xx,\pp] \right\rangle_0,
\end{multline}
where $\xx_0(t)=\xx(t)-\xx(0)$ and $\langle \cdot \rangle_0$ denotes the average over the action of the pure Brownian motion $S_0[\xx,\pp]$. It is shown in \cite{Demery2011} that
\begin{equation}
\left\langle S\ind{int,2}^\mathrm{eff}[\xx,\pp] \right\rangle_0=0;
\end{equation}
thus we only need to compute $\left\langle \xx_0(t)^2 S_\mathrm{int,2}^\mathrm{eff}[\xx,\pp] \right\rangle_0$. The following averages are needed:
\begin{multline}
\left\langle \xx_0(t)^2\pp(t')\ed^{\id\ggamma(Q)\cdot[\xx(t')-\xx(t'')]} \right\rangle_0=-4D_x\chi_{[0,t[}(t')\\ 
\times \mathrm{L}([0,t[\cap [t'',t'[)\ggamma(K)\ed^{-D_x\ggamma(Q)^2 |t'-t''|},
\end{multline}
and
\begin{multline}
\left\langle \xx_0(t)^2\pp(t')\pp(t'')\transp\ed^{\id\ggamma(Q)\cdot[\xx(t')-\xx(t'')]} \right\rangle_0=\\\chi_{[0,t[}(t') \left[-2\chi_{[0,t[}(t'')+4D_x\mathrm{L}([0,t[\cap [t'',t'[)\ggamma(Q)\ggamma(Q)\transp\right]\\
\times\ed^{- D_x \ggamma(Q)^2|t'-t''|}.
\end{multline}
$\chi_I(t)$ is the characteristic function of the interval $I\subset \mathbb{R}$ and $\mathrm{L}(I)$ is its length. We introduce an effective damping for the mode $Q$:
\begin{equation}
\bar\omega(Q)=\omega(Q)+D_x\ggamma(Q)^2.
\end{equation}
Using these expressions and computing the integrals over the interaction times in the long-time limit, we get the effective diffusion coefficient:
\begin{widetext}
\begin{equation}
D\ind{eff}=D_x\\-\frac{1}{d}\int \frac{2D_x \left[\ggamma(Q)\cdot\ff(Q)+T\ggamma(Q)\transp\ggg(Q)\ggamma(Q) \right]-T\tr(\ggg(Q))\bar\omega(Q)}{\bar\omega(Q)^2}\dd Q.
\end{equation}
In our case, it reads
\begin{equation}\label{deff_final}
D\ind{eff}=D_x \left[1-\frac{h^2D_x}{2d}\int \frac{(\kk+\kk')^2 \tilde K(\kk)^2\tilde K(\kk')^2 \left([2\zeta-1]\kappa_\phi \left[\tilde R(\kk)\tilde\Delta(\kk)+\tilde R(\kk')\tilde\Delta(\kk') \right]+D_x[\kk+\kk']^2 \right)}{\tilde\Delta(\kk)\tilde\Delta(\kk') \left(\kappa_\phi \left[\tilde R(\kk)\tilde\Delta(\kk)+\tilde R(\kk'	)\tilde\Delta(\kk') \right] +D_x[\kk+\kk']^2 \right)^2} \frac{\dd\kk\dd\kk'}{(2\pi)^{2d}}\right].
\end{equation}
\end{widetext}



This expression, giving the effective diffusion coefficient of a particle quadratically coupled to a Gaussian fluctuating field outside the adiabatic limit is our main result;
we now discuss some of its features. Firstly, it looks very similar to the expression obtained for a linear coupling in \cite{Demery2011}, that is not surprising since the effective evolution equations, that are the starting point of the calculation, are very similar. The main qualitative difference is that the effective diffusion coefficient $D_x=T\kappa_x$ stands in front of the correction, whereas it is the mobility $\kappa_x$ for a linear coupling. This is a hallmark of fluctuations induced effects, where the temperature plays a central role. It was already observed in \cite{Demery2011a} where the drag coefficient was found to be proportional to the temperature. 

\begin{figure}
\begin{center}
\includegraphics[width=0.95\columnwidth]{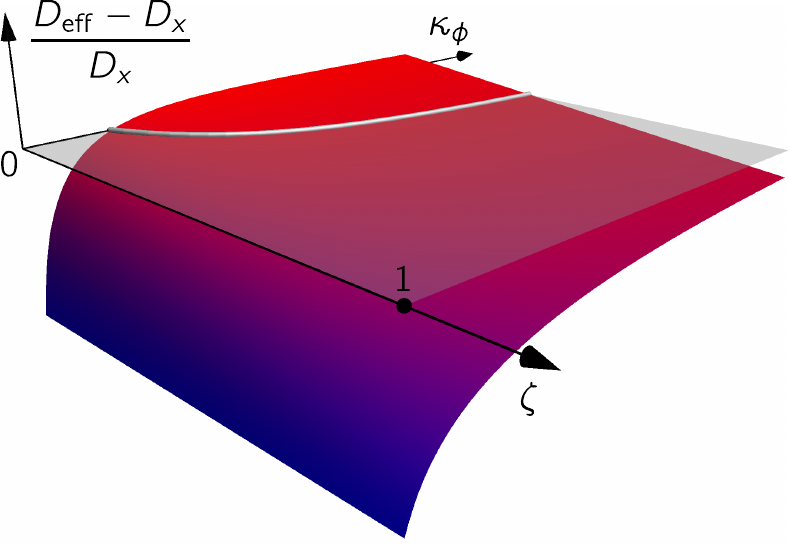}
\vspace{-5mm}
\end{center}
\caption{(Color online) Predicted effective diffusion coefficient as a function of the feedback parameter $\zeta$ and the field evolution rate $\kappa_\phi$, for the parameters used in the numerical simulations. The grey plane indicates the value 0 and the white line is the corresponding contour. This graph shows that the diffusion coefficient is increased for a passive particle in a fast field, and decreased otherwise. }
\label{diag_deff}
\end{figure}

Secondly, the feedback parameter $\zeta$ plays the same role as in the linear coupling case, and the consequences are thus the same: the diffusion coefficient of an active particle is always reduced by its coupling to the field, whereas for a passive particle it is reduced in a slow field and increased in a fast field. This behavior is represented on FIG.~\ref{diag_deff}, where the correction to the diffusion coefficient is plotted against the parameters $\zeta$ and $\kappa_\phi$.

\section{Numerical simulations}\label{sec_num_sim}

We perform numerical simulations on a simple system, in dimension $d=1$ and with a finite number of Fourier modes. All the parameters are kept constant except $\zeta$, that is used to investigate active and passive diffusion, $h$, that allows to test the validity domain of the weak coupling approximation used for the predictions, and $\kappa_\phi$, that allows to vary the field evolution rate and thus interpolate between the quenched field and adiabatic limits.

The fixed parameters and operators are set to $T=1$, $\kappa_x=1$, $\tilde K(k)=\tilde R(k)=1$, $\tilde\Delta(k)=k^2+1$. We use the method introduced in \cite{Lin2004} to simulate a Gaussian field with a finite number of Fourier modes, albeit in dimension $d=1$ here. 
The field reads
\begin{equation}
\phi(y,t)=a_0(t)+\sum_{k=1}^N \left[a_k(t)\cos(ky)+b_k(t)\sin(ky) \right].
\end{equation}
To determine the evolution of the coefficients $a_k(t)$ and $b_k(t)$, we insert this decomposition in the evolution equation \eqref{dyna_champ}, then multiply it by $\cos(ky)$ or $\sin(ky)$ and integrate over $[0,2\pi]$. We get
\begin{align}
\dot a_0(t) & =-\kappa_\phi \left[\tilde\Delta(0)a_0(t)+\frac{\zeta h}{2\pi} \phi(x(t),t) \right]+\sqrt{\kappa_\phi}\xi_0(t)\\
\dot a_k(t) & =-\kappa_\phi \left[\tilde\Delta(k)a_k(t)+\frac{\zeta h}{\pi} \phi(x(t),t) \cos(kx(t)) \right]+\sqrt{\kappa_\phi}\xi_k(t)\\
\dot b_k(t) & =-\kappa_\phi \left[\tilde\Delta(k)b_k(t)+\frac{\zeta h}{\pi} \phi(x(t),t) \sin(kx(t)) \right]+\sqrt{\kappa_\phi}\xi'_k(t)
\end{align}
where the noises have the following correlators
\begin{align}
\left\langle \xi_k(t)\xi_{k'}(t') \right\rangle&= \frac{2T}{\pi}\delta(t-t')\delta_{k-k'}\quad \mathrm{if}\quad k\neq 0,\\
\left\langle \xi_0(t)\xi_0(t') \right\rangle&= \frac{T}{\pi}\delta(t-t'),\\
\left\langle \xi'_k(t)\xi'_{k'}(t') \right\rangle&= \frac{2T}{\pi}\delta(t-t')\delta_{k-k'},\\
\left\langle \xi_k(t)\xi'_{k'}(t') \right\rangle&= 0.
\end{align}

In the simulations, we let the system evolve for a long time $\tau\gg \kappa_x^{-1},\kappa_\phi^{-1}$, and we measure the particle position at regular intervals. We repeat this simulation a large number of times (around $10^5$) and, using these measurements, we compute $\left\langle x(t)^2 \right\rangle$, where $t$ is the measurement time. We then plot this function of $t$ and a linear fit gives the value of the effective diffusion constant.

For a finite number of modes, the integral is replaced by a sum in the effective diffusion coefficient, and the simulations are compared to the following formula
\begin{align}
&D\ind{eff}= 1\\
&-\frac{h^2}{8\pi^2}\sum_{k,k'=-N}^N \frac{(k+k')^2  \left([2\zeta-1]\kappa_\phi \left[\tilde\Delta(k)+\tilde\Delta(k') \right]+[k+k']^2 \right)}{\tilde\Delta(k)\tilde\Delta(k') \left(\kappa_\phi \left[\tilde\Delta(k)+\tilde\Delta(k') \right] +[k+k']^2 \right)^2}.\nonumber
\end{align}

\begin{figure}
\begin{center}
\includegraphics[width=.95\columnwidth]{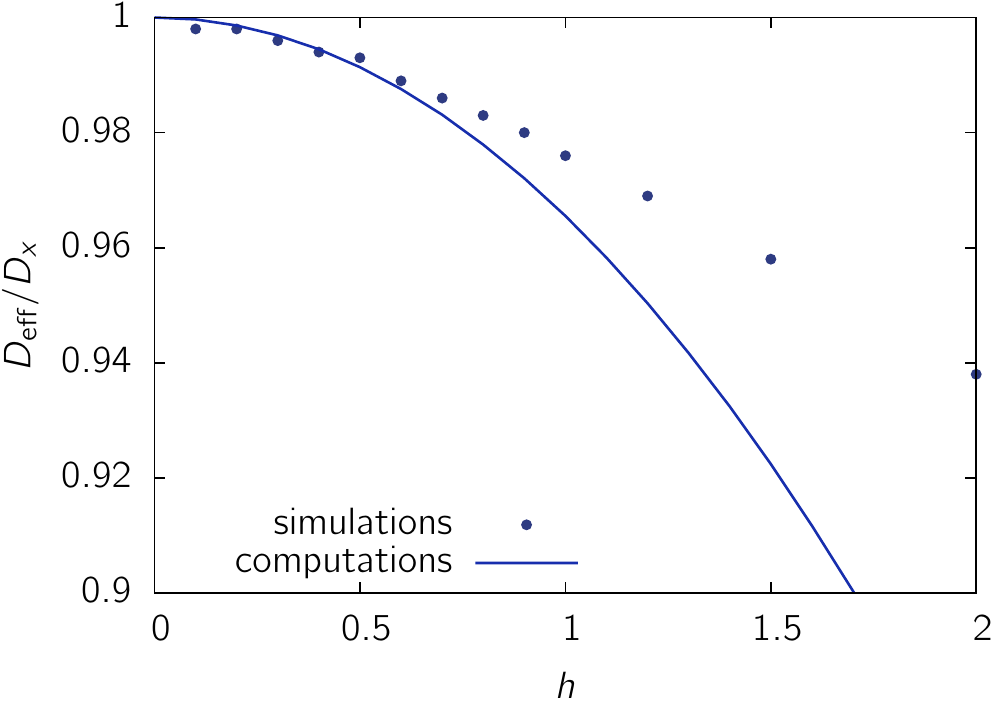}
\vspace{-5mm}
\end{center}
\caption{(Color online) Effective diffusion coefficient of an active particle as a function of the coupling parameter; simulations (circles) and analytical result (continuous line).}
\label{valid_act}
\end{figure}

\begin{figure}
\begin{center}
\includegraphics[width=.95\columnwidth]{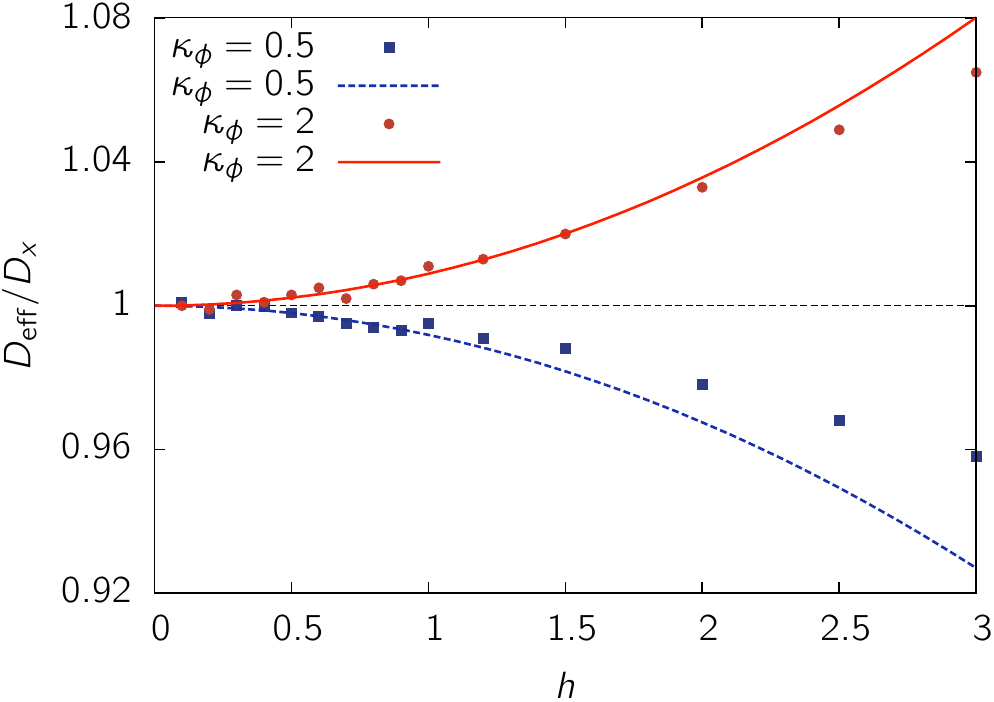}
\vspace{-5mm}
\end{center}
\caption{(Color online) Effective diffusion coefficient of an passive particle as a function of the coupling parameter, for two different field evolution rates; simulations (circles and squares) and analytical result (lines).}
\label{valid_pass}
\end{figure}

The validity domain is questioned in FIG.~\ref{valid_act} for the active case and in FIG.~\ref{valid_pass} for the passive case. For an active particle, it appears that the validity domain is very small: the small coupling approximation holds for $h\lesssim 0.7$, where the correction is only $1\%$ of the bare diffusion coefficient. The situation is different for a passive particle, especially in a fast field ($\kappa_\phi=2$), where the approximation holds for $h\lesssim 2$, where the correction is close to $5\%$. The validity domain is smaller for a slow field ($\kappa_\phi=0.5$). We can conclude that our prediction is better when the diffusion coefficient is increased than when it is decreased; in the last case it rapidly underestimates the diffusion coefficient.

\begin{figure}
\begin{center}
\includegraphics[width=.95\columnwidth]{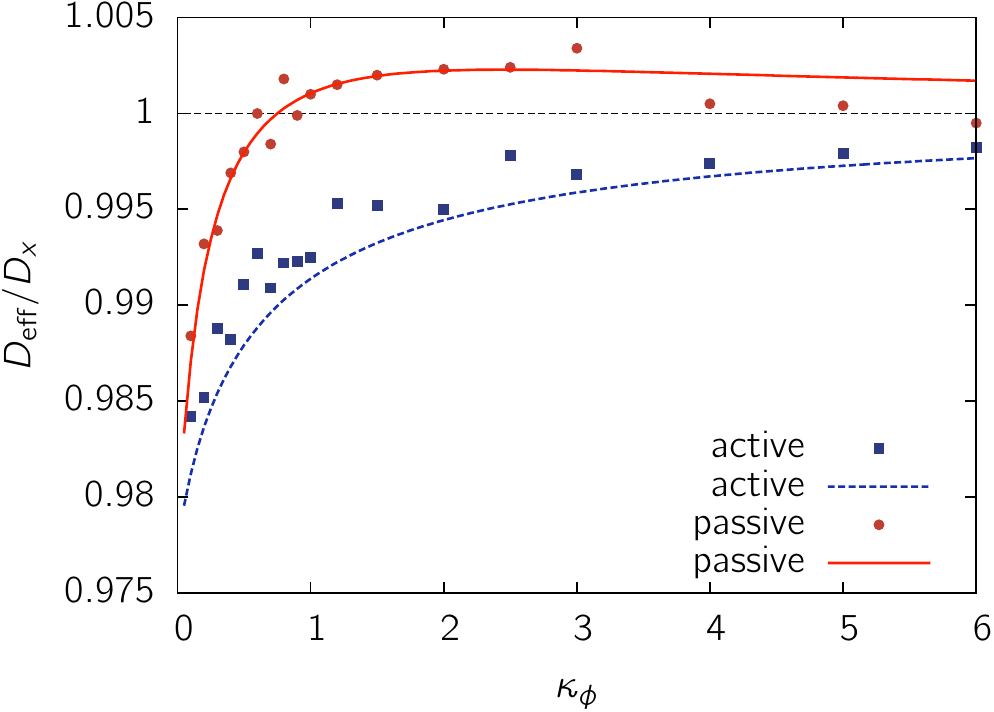}
\vspace{-5mm}
\end{center}
\caption{(Color online) Effective diffusion coefficient as a function of the field evolution rate for a passive and an active particles, with a coupling $h=0.5$; simulations (circles and squares) and analytical result (lines).}
\label{deff_h05}
\end{figure}

\begin{figure}
\begin{center}
\includegraphics[width=.95\columnwidth]{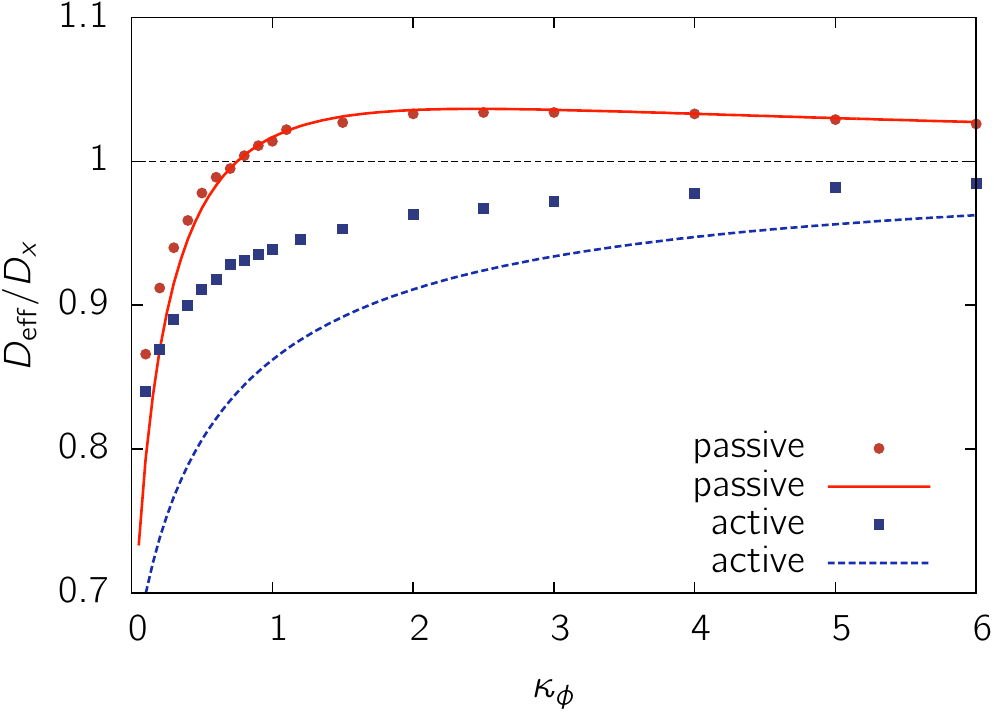}
\vspace{-5mm}
\end{center}
\caption{(Color online) Effective diffusion coefficient as a function of the field evolution rate for a passive and an active particles, with a coupling $h=2$; simulations (circles and squares) and analytical result (lines).}
\label{deff_h2}
\end{figure}

The effect of the field evolution rate is studied in FIG.~\ref{deff_h05} for a small coupling and in FIG.~\ref{deff_h2} for a high coupling. In the first case, the computed diffusion coefficient is close to the one obtained in the simulations. We notice important variations on simulation results; they are due to the fact that the correction to the bare diffusion coefficient is very small, and thus needs a lot of computation time to be determined precisely. At high coupling, the agreement between theory and simulations is not good for an active particle, but the general tendency is correct. On the other hand, the passive calculation appears to be very accurate and fully reproduces the simulations results. Notably, the crossover between a decrease and an augmentation of the diffusion coefficient when the field becomes faster is well captured by the analytic calculation.

As a conclusion, our predictions are precise in a relatively small domain of validity, but they remain qualitatively good at higher coupling, and thus provide a useful understanding of the phenomenon studied here.

\section{Discussion on anomalous diffusion}\label{sec_anomalous}

The study presented here focuses on the computation of small corrections of the effective diffusion constant in the case where the diffusion remains normal. 
However, as explained in \cite{Demery2011,Bouchaud1990} an infrared (small $k$) divergence in the integral invoked in the expression of the diffusion constant indicates a fertile ground for anomalous diffusion. We now discuss the infrared divergence of the integral appearing in (\ref{deff_final}), in a way similar to the one used in \cite{Demery2011}: we start by defining the small $k$ behavior of the operators,
\begin{align}
\tilde\Delta(k) & \underset{k\rightarrow 0}{\sim} k^\delta,\\
\tilde K(k) & \underset{k\rightarrow 0}{\sim} k^\kappa,\\
\tilde R(k) & \underset{k\rightarrow 0}{\sim} k^\rho.
\end{align}
The integral in (\ref{deff_final}) diverges at low $k$ if
\begin{equation}
\mathrm{deg}_0=4\kappa-2\delta-\mathrm{min}(\rho+\delta,2)+2d+2\leq 0,
\end{equation}
where $\mathrm{deg}_0$ is the \emph{degree} of the integral, i.e. its dimension in $\kk$.
In the two examples presented in section \ref{sec_model}, the operator $\Delta$ is of the form $\tilde\Delta(k)=k^\delta \left(k^2+m^2 \right)$. When the field is critical, $m=0$ and $\delta'=\delta+2$ should be used instead of $\delta$ in the previous formula. We now apply this formula to our examples.

The following table gives, for the cases of a protein coupled to membrane curvature or composition, the value of the exponents and the degree $\mathrm{deg}_0$ of the integral (\ref{deff_final}).
\begin{center}
\begin{tabular}{|c|c|c|}
\hline & curvature & composition \\ \hline
parameters & $\delta=2$, $\kappa=2$, $\rho=-1$ & $\delta=0$, $\kappa=0$, $\rho=2$ \\ \hline
non critical & $\mathrm{deg}_0=9$ & $\mathrm{deg}_0=4$ \\ \hline
critical & $\mathrm{deg}_0=4$ & $\mathrm{deg}_0=0$ \\ \hline
\end{tabular}
\end{center}
It appears that a coupling to membrane composition at the miscibility transition can lead to anomalous diffusion. Since the proteins are active ($\zeta=1$), the correction to the diffusion constant is negative and the anomalous diffusion should be subdiffusive. We recall that this statement should be taken carefully: we only say that this system is \emph{likely} to be subdiffusive, we do not assert that it is. For instance, this divergence could also stand for a correction proportional to $h^\alpha$, with $\alpha<2$. The degree 0 of the integral implies a logarithmic divergence, i.e. a correction of the form $\log(l)$ when the system is close to criticality. In this configuration, the temperature dependence of the diffusion coefficient is dominated by the temperature dependence of the correlation length (\ref{mass_temp}).

We now rapidly discuss the case of ultraviolet (large $k$) divergence. An ultraviolet divergence should be regularized with a microscopic cut-off length, for instance the protein size; in this case the diffusion coefficient thus depends on the protein radius, an effect that was explored experimentally in \cite{Gambin2006} and discussed in \cite{Demery2010,Demery2011a}. The divergence criterion is here
\begin{equation}
\mathrm{deg}_\infty=4\kappa-2\delta-\mathrm{max}(\rho+\delta,2)+2d+2\geq 0,
\end{equation}
where the exponents are defined in the limit $k\rightarrow\infty$; this analysis does not depend on the correlation length. We find that $\mathrm{deg}_\infty=3$ for a coupling to the curvature and $\mathrm{deg}_\infty=-2$ for a coupling to the composition. In the first case, regularizing the divergence with the protein radius $a$ gives a correction proportional to $a^3$.

\section{Conclusion}\label{sec_conclu}

We analyzed the diffusion of a particle quadratically coupled to a fluctuating field, that avoids regions where the field fluctuates and may reduce the field fluctuations, in which case it is called active; if it does not affect the field it is called passive. This study completes the one developed in \cite{Demery2011} for a linear coupling between the field and the particle and they set together a general model for diffusion in a fluctuating environment. The environment is modeled by a Gaussian field obeying overdamped Langevin dynamics. The particle-field interaction can take two forms: if the particle breaks the symmetry of the field and prefers one specific sign, it is modeled by a linear interaction; if it does not break the symmetry of the field and avoids field fluctuations, it is represented by a quadratic interaction. 
Both types are present in the general case, an example being the protein coupled to membrane curvature discussed in section \ref{ssec_model_examples}.
In the limit of a weak interaction, both interaction types contribute to the full correction to the effective diffusion coefficient; alternatively a preliminary study may determine the dominant contribution.

We focused on the computation of the effective diffusion coefficient, assuming normal diffusion; our method allows to deal with equilibrium as well as out of equilibrium systems (that is the case if the environment is not affected by the particle or subject to non thermal forces \cite{Douglass2012,Sens2011}). Our conclusion is that for both types of coupling, the diffusion coefficient is always reduced for an active particle, whereas it can be reduced or increased for a passive particle if the field is slow or fast, respectively; this is summarized in FIG.~\ref{diag_deff}. 

When the correction to the bare diffusion coefficient diverges, it may be interpreted as the onset of anomalous diffusion; depending on the sign of the correction, it indicates subdiffusion or superdiffusion. For instance, for membrane proteins, we have shown that subdiffusion may occur when the protein is coupled to the membrane composition at the critical point.
This output of our theory is of major importance regarding the numerous examples of anomalous diffusion found in experiments \cite{Tabei2013,Douglass2012}; however, it gives little information on the diffusion and notably on the exponent. This limitation comes from the fact that we use a perturbative expansion around normal diffusion and compute the prefactor, i.e. the diffusion coefficient. Further work is thus needed to be able to deal with environment induced anomalous diffusion; 
a necessary step in this direction is to go beyond perturbative results.







\section{Acknowledgments}\label{}

I acknowledge funding from the Institut des Systèmes Complexes de Paris Île de France. 

\appendix

\section{Functional operators in real and Fourier space}\label{ap_operators}

In this appendix, we define our notations and recall some basic properties of functional operators. We start in real space, and then see how it transposes to Fourier space, that is used a lot in this article. All the operators considered here are real. 

For two functions $f(\yy)$ and $g(\yy)$ and two operators $A(\yy,\yy')$ and $B(\yy,\yy')$, the scalar product of $f$ and $g$, the action of $A$ on $f$, the product of $A$ and $B$ and the trace of the operator $A$ are respectively defined by
\begin{align}
f\cdot g & = \int f(\yy)g(\yy)\dd \yy,\label{prod_scal}\\
(Af)(\yy) & = \int A(\yy,\yy')f(\yy')\dd \yy',\\
(AB)(\yy,\yy') & = \int A(\yy,\yy'')B(\yy'',\yy')\dd \yy'',\\
\tr(A) &=\int A(\yy,\yy)\dd\yy.\label{trace}
\end{align}

The adjoint $A^\dagger$ of the operator $A$ is defined by the equality $f\cdot(Ag)=\left(A^\dagger f \right)\cdot g$, that holds for all functions $f$ and $g$; it is straightforward to show that
\begin{equation}
A^\dagger(\yy,\yy')=A(\yy',\yy).
\end{equation}
$A$ is symmetric or self-adjoint if $A^\dagger=A$, or 
\begin{equation}
A(\yy,\yy')=A(\yy',\yy).
\end{equation}
An operator $A$ is invariant by translation if there exists a function $a$ such that
\begin{equation}
A(\yy,\yy')=a(\yy-\yy').
\end{equation}
Such an operator is isotropic if it only depends on the distance between $\yy$ and $\yy'$:
\begin{equation}
A(\yy,\yy')=a\left(|\yy-\yy'|\right).
\end{equation}

We now switch to the Fourier space, with the Fourier transform defined by
\begin{align}
f(\yy) & = \int \ed^{\id\kk\cdot\yy} \tilde f(\kk)\frac{\dd\kk}{(2\pi)^d},\\
A(\yy,\yy') & = \int \ed^{\id(\kk\cdot\yy+\kk'\cdot\yy')} \tilde A(\kk,\kk')\frac{\dd\kk\dd\kk'}{(2\pi)^{2d}},
\end{align}
for a function and an operator, respectively. This definition allows us to translate (\ref{prod_scal}-\ref{trace}) into Fourier space:
\begin{align}
f\cdot g & = \int \tilde f(-\kk)g(\kk)\frac{\dd\kk}{(2\pi)^d},\\
\widetilde{Af}(\kk) & = \int \tilde A(\kk,-\kk')\tilde f(\kk')\frac{\dd\kk'}{(2\pi)^d},\\
\widetilde{AB}(\kk,\kk') & = \int \tilde A(\kk,-\kk'')B(\kk'',\kk')\frac{\dd\kk''}{(2\pi)^d},\\
\tr(A) &=\int \tilde A(\kk,-\kk)\frac{\dd\kk}{(2\pi)^d}.
\end{align}
The adjoint of the operator $A$ is defined in the same way as in real space:
\begin{equation}
\widetilde{A^\dagger}(\kk,\kk') =\tilde A(\kk',\kk).
\end{equation}
The Fourier transform of the translation-invariant operator $A(\yy,\yy')=a(\yy-\yy')$ reads
\begin{equation}
\tilde A(\kk,\kk')=(2\pi)^d\tilde a(\kk)\delta(\kk+\kk').
\end{equation}
Moreover, if $A$ is isotropic, its Fourier transform only depends on the norm $|\kk|$: $\tilde a(\kk)=\tilde a(|\kk|)$.

In this article, we do not use a different notation for the one-variable function associated to a translation-invariant operator: the number of variables indicates if we refer to the operator or to its associated function. For instance, we will use $\tilde \Delta(\kk,\kk')=(2\pi)^d\tilde \Delta(\kk)\delta(\kk+\kk')$.

\section{Field correlation functions}\label{ap_correl_champ}

In this section, we derive the free field correlation functions $\langle \psi \psi\rangle$, $\langle \phi\psi \rangle$ and $\langle \phi\phi \rangle$ from the free field action (\ref{act_champ_libre}). Equations on averages can be obtained from the action with the Schwinger-Dyson formula \cite{Dyson1949b}:
\begin{equation}
\left\langle \frac{\delta A[\phi,\psi]}{\delta\phi(\yy,t)} \right\rangle=\left\langle A[\phi,\psi]\frac{\delta S_0^\phi[\phi,\psi]}{\delta\phi(\yy,t)} \right\rangle,
\end{equation}
where $A$ is a functional operator. The same equation holds if the derivative is taken with respect to $\psi(\yy,t)$. Using $A=1$, we derive two equations that give the averages
\begin{align}
\langle \phi(\yy,t) \rangle&=0,\\
\langle \psi(\yy,t) \rangle&=0.
\end{align}

With $A[\phi,\psi]=\psi(\yy,t)$ and derivating with respect to $\phi(\yy',t')$, we get 
\begin{equation}
0=\left\langle \psi(\yy,t) \left[\dot\psi(\yy',t')-\kappa_\psi(R\Delta\psi)(\yy',t') \right] \right\rangle,
\end{equation}
that leads to
\begin{equation}
C_{\psi\psi}(\yy,\yy',t,t')=\left\langle \psi(\yy,t)\psi(\yy',t') \right\rangle=0.
\end{equation}
Setting $A[\phi,\psi]=\phi(\yy,t)$ and derivating with respect to $\phi(\yy',t')$ gives
\begin{equation}
\delta(\yy-\yy')\delta(t-t')=\id \left\langle \phi(\yy,t) \left[\dot\psi(\yy',t')-\kappa_\phi(R\Delta\psi)(\yy',t') \right] \right\rangle.
\end{equation}
This equation has the solution
\begin{align}
C_{\phi\psi}(\yy,\yy',t,t')&= \left\langle \phi(\yy,t)\psi(\yy',t') \right\rangle \\
&= \id\left[\ed^{-\kappa_\phi (t-t')R\Delta}\right](\yy-\yy')\theta(t-t'). \label{correl_champ_phipsi}
\end{align}
Finally, keeping the same operator $A$ and derivating with respect to $\psi(\yy',t')$, we get
\begin{multline}
\id \left\langle \phi(\yy,t) \left[\dot\phi(\yy',t')+\kappa_\phi(R\Delta\phi)(\yy',t') \right] \right\rangle=\\2T\kappa_\phi \left\langle \phi(\yy,t)(R\psi)(\yy',t') \right\rangle.
\end{multline}
Using the previous result (\ref{correl_champ_phipsi}), we solve this equation with
\begin{align}
C_{\phi\phi}(\yy,\yy',t,t')&= \left\langle \phi(\yy,t)\phi(\yy',t') \right\rangle \\
&= T\left[\Delta^{-1}\ed^{-\kappa_\phi |t-t'|R\Delta}\right](\yy-\yy').
\end{align}

\bibliographystyle{apsrev}


\end{document}